\documentclass[preprint,superscriptaddress,nofootinbib]{revtex4-1}
\usepackage{graphicx}
\usepackage{dcolumn}
\usepackage{bm}
\usepackage{amsmath}
\usepackage{amsfonts} 
\usepackage{latexsym}
\usepackage{bbm}
\usepackage{color}
\usepackage{amssymb}
\usepackage{amsthm}
\usepackage{epsf}
\usepackage{epsfig}
\usepackage{caption3}
\usepackage{appendix}
\usepackage{cases}
\usepackage{subfigure}
\usepackage{multirow}
\usepackage{amsthm,amsmath,amssymb}
\usepackage{mathrsfs}
\makeatletter

\newcommand{\Rmnum}[1]{\expandafter\@slowromancap\romannumeral #1@}
\makeatother

\newcommand{\tc}{\textcolor{black}}
\newcommand{\rc}{\textcolor{black}}
\begin{document}

\title{Implication of nano-Hertz stochastic gravitational wave on
dynamical dark matter through a \rc{dark} first-order phase transition}

\author{Siyu Jiang}

\author{Aidi Yang}

\author{Jiucheng Ma}

\author{Fa Peng Huang}%
\email{Corresponding author. huangfp8@sysu.edu.cn}

\affiliation{MOE Key Laboratory of TianQin Mission, TianQin Research Center for
	Gravitational Physics \& School of Physics and Astronomy, Frontiers
	Science Center for TianQin, Gravitational Wave Research Center of
	CNSA, Sun Yat-sen University (Zhuhai Campus), Zhuhai 519082, China}

\bigskip
	
	
\begin{abstract}
For the first time, the expected stochastic gravitational wave background is probably
discovered after observing the Hellings Downs correlation curve by several pulsar timing array (PTA) collaborations around the globe including NANOGrav, European PTA, Parkes PTA, and Chinese PTA. 
These new observations can help to explore or constrain the dark matter formation mechanisms in the early universe.
We study the implication of those results on the dynamical dark matter formation mechanisms through a \rc{dark} first-order phase transition in the early universe. Both the Q-ball dark matter and super-cool dark matter are investigated in the  strong super-cooling \rc{dark} phase transition scenario which may give an interpretation of the observed stochastic gravitational wave background.
\end{abstract}

\maketitle

\section{Introduction }	
Recently, various pulsar timing array (PTA) collaborations from NANOGrav, European PTA, Parkes PTA, and Chinese PTA~\cite{NANOGrav:2023gor,Antoniadis:2023ott,Antoniadis:2023zhi,Reardon:2023gzh,Xu:2023wog} have published their most recent findings on the first observation of the leading-order overlap reduction function, namely, the famous Hellings Downs~\cite{Hellings:1983fr} curve which supports the discovery of the stochastic gravitational wave background (SGWB). 
Besides the astrophysical sources of SGWB (such as the populated supermassive black hole binary), there are abundant cosmological sources of SGWB produced in many important processes in the early universe. With the discovery of SGWB, it becomes realistic to explore the new physics beyond the standard model (SM) and our early Universe by the PTA measurements or nano-Hertz SGWB~\cite{NANOGrav:2023hvm,Bringmann:2023opz,Madge:2023cak,Chiang:2020aui,Gouttenoire:2023ftk,Cai:2023dls,Broadhurst:2023tus,Bai:2023cqj,Shen:2023pan,Yang:2023aak,Fujikura:2023lkn,Oikonomou:2023qfz,Megias:2023kiy,Athron:2023xlk}. \tc{Based on the new data of NANOGrav, Ref.~\cite{NANOGrav:2023hvm} comprehensively studied various new physics models and compared them by Bayesian analysis. These studies include the ultralight dark matter (DM), cosmological phase transition, cosmic strings, cosmic inflation, scalar-induced gravitational waves, domain walls, etc~\cite{NANOGrav:2023hvm,Bringmann:2023opz,Madge:2023cak,Chiang:2020aui,Gouttenoire:2023ftk,Cai:2023dls,Broadhurst:2023tus,Bai:2023cqj,Shen:2023pan,Yang:2023aak,Fujikura:2023lkn,Oikonomou:2023qfz,Megias:2023kiy,Athron:2023xlk}.} In this work, we focus on the SGWB produced in the DM formation process through a strong \rc{dark} first-order phase transition (FOPT) in the early Universe. 

\rc{In contrast with the electroweak phase transition where the Higgs field is the phase transition field, the phase transition field of the dark FOPT is usually the new scalar field in the dark sector. Thus, compared with the electroweak FOPT, the dark FOPT is less constrained
by colliders and other experiments.
The FOPT can be hidden in dark sector where the dark particles weakly couple  with the SM. 
The dark FOPT can easily happen at temperatures lower than the electroweak scale. The dark FOPT can not only produce low frequency SGWB~\cite{Schwaller:2015tja,Addazi:2017gpt,Addazi:2020zcj} but also explain the DM and baryon asymmetry~\cite{Hall:2019ank,Hall:2019rld,Nakagawa:2022wwm,Ahmadvand:2021vxs,Baker:2021zsf}.  Because of this, phase-transition gravitational waves have become an important way to detect the dark FOPT models.}

Motivated by the absence of the expected DM signals at colliders and DM direct search, new DM production mechanisms and new detection approaches attract lots of intriguing studies. Among them, the dynamical DM formation mechanism through a strong FOPT provides a natural new mechanism for
a wide DM mass range which strongly depends on the phase transition dynamics, such as the bubble wall velocity. Meanwhile, this type of dynamical mechanism
could be directly detected by the associated phase transition SGWB with the DM production~\footnote{The anisotropy of the SGWB through future SKA can provide more information~\cite{Li:2021iva} of the early universe.}. 

The peak frequency of SGWB produced by a FOPT is proportional to the phase transition temperature, the inverse duration of phase transition $\beta/H$, and is inversely proportional to bubble wall velocity. Generally, low-frequency gravitational wave spectra require a low-temperature phase transition. However, the FOPT temperature is strongly constrained by Big Bang nucleosynthesis (BBN). In order to avoid this, the nano-Hertz SGWB prefers a slow phase transition, i.e., small $\beta/H$ and a large bubble wall velocity~\cite{Bringmann:2023opz}. This can be naturally realized in super-cooling phase transition\cite{Guth:1981uk,Wang:2020jrd}. And in the super-cooling phase transition, as the DM particles can gain much larger mass compared with the temperature, the DM relic density can be produced by some more interesting dynamical mechanisms such as the Q-ball DM~\cite{Huang:2017kzu,Krylov:2013qe}, the filtered DM~\cite{Baker:2019ndr,Chway:2019kft,Jiang:2023nkj} and the super-cool DM~\cite{Hambye:2018qjv}. \tc{These dynamical DM mechanisms are specified by the DM penetration behavior into the bubble which depends on the DM mass and bubble wall velocity. When the penetration rate is approximately zero, or in other words, most DM particles are trapped in the false vacuum, we can get Q-ball or Fermi-ball DM which depends on whether the constituent particles are bosons or fermions. For the filtered DM, the suppressed DM number density penetrating into the bubble is naturally out-of-equilibrium and can easily fulfil the DM relic density.} On the other hand, the super-cooling phase transition could induce a period of inflation which can naturally dilute the DM relic density even though the penetration rate is high enough. This is the so-called super-cool DM mechanism.

 In this work, we focus on two types of dynamical DM mechanisms which are based on the phase transition dynamics during a \rc{super-cooling dark} FOPT: the Q-ball or Fermi-ball DM and the super-cool DM. \tc{These scenarios are differentiated by different penetration rates.  We aim to correlate these scenarios with different penetration rates and provide comparative SGWB spectra for each DM mechanism, aligning our findings with signals observed by NANOGrav.} In Sec.~\ref{PTD}, we briefly review the dynamics of phase transitions. The toy model which introduces a strong super-cooling phase transition is reviewed in Sec.~\ref{conformal} and the penetration rate of DM in the dark FOPT process is discussed in Sec.~\ref{PE}. In Sec.~\ref{DC} we mainly consider the Q-ball (Fermi-ball) DM and super-cool DM. \tc{We also show the SGWB spectra for these two mechanisms.} A concise conclusion is given in Sec.~\ref{con}.

\section{Phase Transition dynamics}\label{PTD}
In the early Universe, a FOPT may happen 
through bubble nucleation 
when there is a specified potential barrier between the false vacuum and the true vacuum.
As an example, before the phase transition, the global minimum of the scalar effective potential $U(\phi,T)$ is located at $\langle\phi\rangle=0$. At the critical temperature $T_c$, two degenerate minima occur and are separated by a barrier.
When the temperature drops below $T_c$, the phase at $\langle\phi\rangle=v(T)$ becomes the new global minimum, also known as the true vacuum. This is the point at which the universe begins its decay towards the true vacuum. When the temperature reaches the nucleation temperature $T_n$, bubbles begin to nucleate.
The nucleation rate is given by 
\begin{equation}
	\Gamma(T) \approx T^4\left(\frac{S_3(T)}{2\pi T}\right)^{3/2} e^{-S_3(T) / T},
\end{equation}
where $S_3(T)$ being the action of the $O(3)$ symmetric bounce solution.
We typically adopt the definition of the nucleation temperature when the nucleation rate initially matches the Hubble rate,
\begin{equation}\label{Tn}
	\Gamma\left(T_n\right) H^{-4}\left(T_n\right) \approx 1 ,
\end{equation}
where $H(T)$ is the Hubble rate,
\begin{equation}
	H^2(T)=\frac{8 \pi}{3 M_{\mathrm{pl}}^2}\left(\frac{\pi^2}{30} g_\star T^4+\Delta U(T)\right) .
\end{equation}
where $M_{\mathrm{pl}}=1.22 \times 10^{19} \,\,\mathrm{GeV}$  is the Planck mass, $g_\star$ is the number of relativistic degrees of freedom (dof). $\Delta U(T)$ is the potential energy difference between true and false vacuum.
\begin{equation}\label{4}
	\Delta U(T)=U(0, T)-U(v(T), T)\,\,.
\end{equation}
The energy release $\Delta U(T)$ in Eq.~\eqref{4} is generally much smaller than the universe's radiation energy $\pi^2g_\star T^4/30$ in ordinary phase transitions. However, in our scenario, a super-cooling FOPT~\cite{Guth:1981uk} is characterized by a delayed nucleation onset compared to the critical temperature $T_c$ \cite{Hong:2020est}. This results in a large releasing energy comparable to the radiation energy. 

Initially, the formed bubbles are minuscule but expand rapidly due to the energy difference between the inside and outside of the bubble. Consequently, the volume of the old phase diminishes with time. \tc{This change can be quantified by the probability $p(T)$ of finding a point in the false vacuum~\cite{Turner:1992tz,Megevand:2016lpr,Kobakhidze:2017mru},}
\begin{eqnarray}\label{Tp}
p(T)=e^{-I(T)}\,\,,
\end{eqnarray}
where $I(T)$ is the fraction of vacuum converted to the true vacuum,
\begin{eqnarray}\label{IT}
	I(T)=\frac{4 \pi}{3} \int_T^{T_c} d T^{\prime} \frac{\Gamma\left(T^{\prime}\right)}{T^{\prime 4} H\left(T^{\prime}\right)}\left[\int_T^{T^{\prime}} d \tilde{T} \frac{v_w}{H(\tilde{T})}\right]^3\,\,.
\end{eqnarray}
It is expected that $p(T)$ approaches zero as the FOPT proceeds and finishes.
The percolation temperature, $T_p$, is defined as the temperature at which bubbles form an infinite connected cluster, with $p(T_p) = 0.71$~\cite{Turner:1992tz}, corresponding to $I(T_p) = 0.34$. The percolation temperature is actually the temperature at which the SGWB from a FOPT is produced~\cite{Megevand:2016lpr,Kobakhidze:2017mru,Ellis:2020awk,Ellis:2018mja}. After percolation, the fraction of the old phase continues to decrease. 

In a super-cooling phase transition, the phase-transition strength $\alpha$ can be expressed by the ratio of the vacuum to the radiation energy:
\begin{equation}
    \alpha\approx\frac{\Delta U(T)}{\rho_{\mathrm{rad}}}\,\,,
\end{equation}
where $\rho_{\mathrm{rad}}=\pi^{2}g_{\star}T^{4}/30$ represents radiation energy density.
 
 The inverse time duration $\beta$ at fixed temperature $T_i$ is generally defined as
\begin{equation}
\beta=H(T) \left.T \frac{d}{d T}\left(\frac{S_3}{T}\right)\right|_{T=T_i}\,\,.
\end{equation}
In the subsequent text, the symbols  $\alpha_p=\alpha(T_p)$ and $H_p=H(T_p)$ will be used to denote $\alpha$ and $H$ at the percolation temperature individually.
\section{strong super-cooling phase transition model}\label{conformal}
The dynamical DM formation mechanisms, usually favour a strong super-cooling phase transition model in which the DM particles get a large mass compared to the phase transition temperature.

\begin{figure}[htbp]
	\begin{center}
		\includegraphics[width=0.7\linewidth]{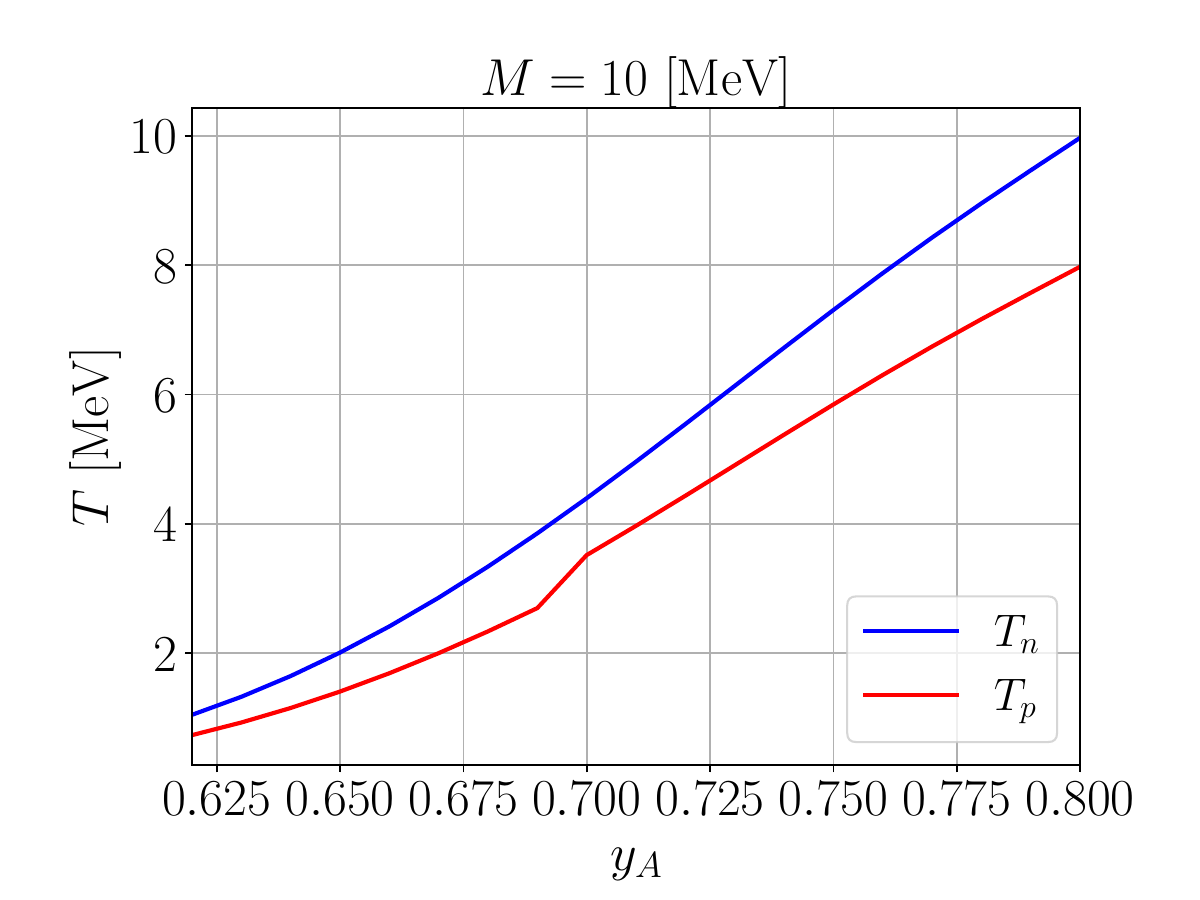}
		\caption{The values of nucleation temperature $ T_n $ and percolation temperature $ T_p $ as function of the $y_A$. We choose $ M = 10~\mathrm{MeV}$.}
		\label{TyA}
	\end{center}
\end{figure}

In this work, we choose the well-studied conformal model. The (classical) scale-invariance or conformal symmetry is realized by preventing any massive parameters in the Lagrangian. \rc{See Refs.~\cite{Kierkla:2022odc,Levi:2022bzt} for details}. \rc{Generally, the classically flat direction of the scalar field is lifted by radiative corrections, the conformal symmetry is therefore radiatively broken. In the mechanism of the radiative symmetry breaking, the one-loop quantum correction should be comparable to the tree-level contribution to the effective potential. One important feature in this scenario is that the potential barrier exists even at zero temperature, so the phase transition strength or the vacuum expectation value over the temperature is usually large. } \tc{In this work, we consider a dark phase transition model~\cite{Levi:2022bzt} where one introduces a complex scalar field $\Phi=(\phi+iG)/\sqrt{2}$. $\phi$ is the order parameter of phase transition and $G$ is the Goldstone boson.}
The conformal potential generally induces a flat direction, along which the leading-order effective potential consists of the Coleman-Weinberg zero-temperature and the thermal one-loop potential (The tree-level potential at zero temperature vanishes.). The former is given by~\cite{Coleman:1973jx}
\begin{equation}
U_{C W}=\sum_i g_i \frac{m_i^4(\phi)}{64 \pi^2}\left[\log \left(\frac{m_i^2(\phi)}{\Lambda_u^2}\right)-c_i\right]\,\,,
\end{equation}
where the sum runs over species with field-dependent masses $m_i^2(\phi)$$.$ $ g_i$ counts the dof for each species, $c_i=5 / 6~(3 / 2)$ for gauge bosons (scalars and fermions), and $\Lambda_u$ is the renormalization scale. The one-loop thermal corrections is
\begin{equation}
U_T=\sum_i \frac{g_i T^4}{2 \pi^2} \int_0^{\infty} d x x^2 \log \left(1-e^{-\sqrt{\left(m_i(\phi) / T\right)^2+x^2}}\right)-\sum_j \frac{g_j T^4}{2 \pi^2} \int_0^{\infty} d x x^2 \log \left(1+e^{-\sqrt{\left(m_j(\phi) / T\right)^2+x^2}}\right).
\end{equation}
$g_i$ and $g_j$ represent the dof for boson and fermion species respectively. In our case, we consider only \tc{gauge boson $A$} with $g_A=3$~\cite{Levi:2022bzt,Madge:2023cak}. \rc{The complex scalar $\phi$ is charged under a U(1) gauge symmetry with coupling strength $y_A$.} $m_A(\phi)=y_A \phi$ is the field-dependent mass of the gauge boson. Then summing $U_{CW}$ and $U_{T}$, at high temperature we have
\begin{equation}
U(\phi, T)=m^2(T) \phi^2-\epsilon(T)\phi^3+\lambda(T)\phi^4\,\,,
\end{equation}
where
\begin{equation}
m^2(T)=g_A \frac{y_A^2 T^2}{24}, \quad \epsilon(T)=g_A \frac{y_A^3 T}{12 \pi}, \quad \lambda(T)=g_A \frac{y_A^4}{32 \pi} \log \left(\frac{T}{M}\right)\,\,,
\end{equation}
and $M \equiv e^{-\frac{1}{3}+\gamma_E} \Lambda_u / 4 \pi$ 
with $\gamma_E$ is the Euler-Mascheroni constant. \tc{Here we neglect the contributions from Daisy resummations. In this work, we focus on the parameter space $y_A \gtrsim 0.6$ which can be seen in Fig.~\ref{TyA}. The Daisy contributions become important for $y_A \lesssim 0.53$~\cite{Levi:2022bzt}, and hence be negligible for
	most of the parameter space considered in this work.}

\tc{With the effective potential in hand, we can determine the nucleation rate by solving the bounce equation.} The bounce action can be parameterized analytically as follows~\cite{Levi:2022bzt}:
\begin{eqnarray}
    \frac{S_3}{T}=
\begin{cases}
\frac{4\pi^3}{27y_{A}^3}\frac{1}{(k-k_c)^2}B_3(k), & k>0\,\,,\\
\frac{3\pi^3}{y_{A}^3}\left(\frac{1+e^{-1/\sqrt{|k|}}}{1+9|k|/2}\right), & k < 0\,\,,
\end{cases}
\end{eqnarray}
where $k(T)=1/6~\mathrm{log}(T/M)$, $k_c=k(T_c)=2/9$ and
\begin{equation}
    B_3(k)=\frac{16}{243}\left[1-38.23(k-\frac{2}{9})+115.26(k-\frac{2}{9})^2+58.07\sqrt{k}(k-\frac{2}{9})^2+229.07k(k-\frac{2}{9})^2\right]\,\,.
\end{equation}
We then use Eqs.~\eqref{Tn} and \eqref{Tp} to calculate the nucleation temperature $T_n$ and percolation temperature $T_p$ respectively. Fig.~\ref{TyA} shows the values of $ T_n $ and $ T_p $ as functions of the $y_A$ with $ M = 10~\mathrm{MeV} $.

\begin{figure}[htbp]
	\centering
	\subfigure{
		\begin{minipage}[t]{0.5\linewidth}
			\centering
			\includegraphics[scale=0.65]{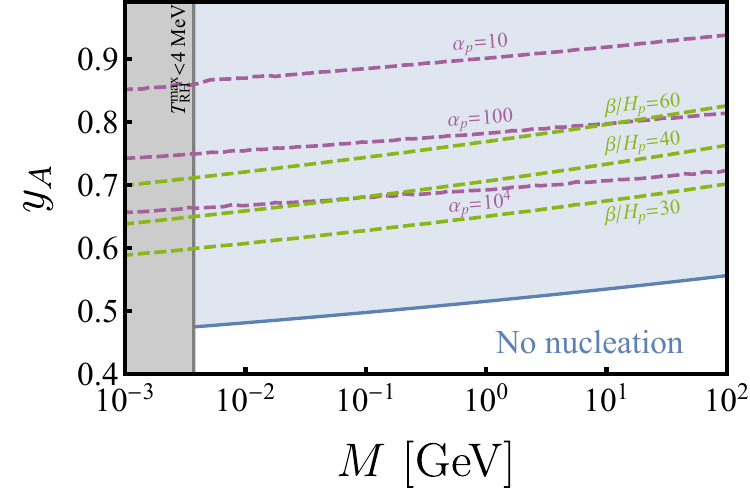}
	\end{minipage}}%
	\subfigure{
		\begin{minipage}[t]{0.5\linewidth}
			\centering
			\includegraphics[scale=0.65]{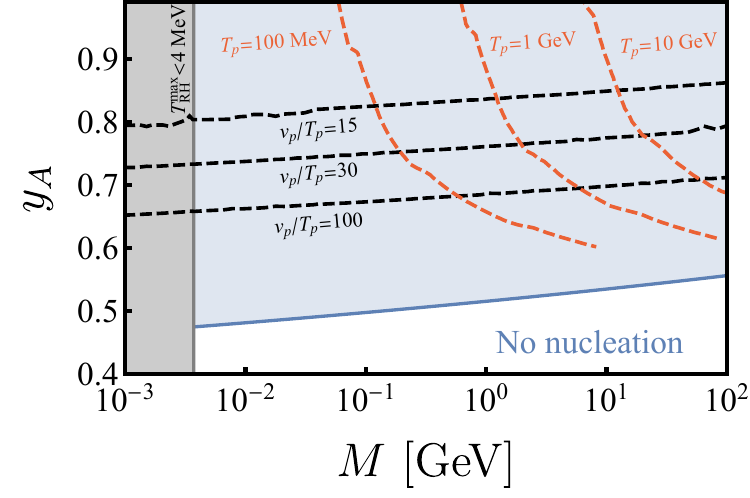}
	\end{minipage}}	
	\caption{{Left: The phase-transition strength $\alpha_p$ and inverse duration $\beta/H_p$ at percolation as functions of $y_A$ and $M$. In the white region, nucleation does not occur.
    Right: The percolation temperature $T_p$ and the wash-out parameter with identical parameters $y_A$ and $M$.}}
      \label{parameters}
\end{figure}

In Fig.~\ref{parameters} we show the phase transition parameters at the percolation temperature: phase-transition strength $\alpha_p$, inverse duration $\beta/H_{p}$,  wash-out parameter $v_p/T_p$ and the percolation temperature $T_p$ in the region of $y_A$ and $M$. Generally we require the $T_{\mathrm{RH}}^{\mathrm{max}}=T_{\mathrm{infl}} \gtrsim 4~\mathrm{MeV}$ in order to evade the constraint of BBN~\cite{deSalas:2015glj} where the $T_{\mathrm{infl}}$ is the temperature of the onset of the inflation. The white region depicts the parameter region where the nucleation cannot be realized.

\section{Penetration rate of dark matter across the bubble wall}\label{PE}
Only particles with enough energy can pass through the bubble wall when they hit the wall. In bubble wall frame, this condition reads~\cite{Baker:2019ndr,Chway:2019kft,Jiang:2023nkj},
\begin{eqnarray}
p_z^w>\sqrt{\Delta m^2},
\end{eqnarray}
where $p_z^w$ is the particle $z$-direction momentum in the bubble wall frame, $\Delta m^2=m_\chi^2-m_0^2$ with $m_\chi=y_\chi v$ being the mass of $\chi$ after phase transition and $m_0$ being the mass in the false vacuum~\cite{Chao:2020adk}. In our work, we set $m_0=0$. $y_\chi$ is the coupling between DM and the scalar field. 

Assuming DM is in thermal equilibrium outside the bubble, then in the bubble wall frame its distribution function follows the Bose-Einstein or Fermi-Dirac distribution
\begin{eqnarray}
f_\chi^{\rm eq}=\frac{1}{e^{ \gamma_{w}\left(|{\mathbf{p}^w}|- v_{w} {p}_z^w\right) / T}\mp 1},
\end{eqnarray}
where $\gamma_{w}=1/\sqrt{1-v_w^2}$ is the Lorentz factor. The particle flux coming from the false vacuum per unit area and unit time is~\cite{Baker:2019ndr,Chway:2019kft},
\begin{eqnarray}\label{flux}
{J}_\chi^w=g_{\mathrm{DM}} \int \frac{d^3 {\mathbf{p}^w}}{(2 \pi)^3} \frac{{p}_z^w}{{E}^w} f_\chi^{\rm eq} \Theta\left({p}_z^w-\sqrt{\Delta m^2}\right) \text {, }
\end{eqnarray}
\tc{where the Heaviside step function $\Theta$ ensures that only particles with sufficient momentum contribute and $g_{\mathrm{DM}}$ is the DM dof.} Then the DM number density inside the bubble $n_\chi^{\text {true }}$ in plasma rest frame can be denoted as 
\begin{eqnarray}\label{nin}
n_\chi^{\text {true }}=\frac{{J}_\chi^w}{ \gamma_{w}  v_{w}} .
\end{eqnarray}
Finally, the DM penetration rate $R_{\rm in}$ follows the form
\begin{eqnarray}
R_{\rm in}=\frac{n_\chi^{\text {true }}}{n_\chi^{\text {false }}},
\end{eqnarray}
where $n_\chi^{\text {false }}$ is the DM number density outside the bubble wall. For $m_0=0$, we can integrate Eq.~\eqref{flux} and get
\begin{eqnarray}\label{nchifilter}
n_\chi^{\text {true }} \simeq \frac{g_{\mathrm{DM}} T_{}^3}{\gamma_{w}  v_{w}}\left(\frac{ \gamma_{w}\left(1-  v_{w}\right) m_\chi / T_{}+1}{4 \pi^2 \gamma_{w}^3\left(1-  v_{w}\right)^2}\right) e^{-\frac{ \gamma_{w}\left(1- v_{w}\right) m_\chi}{T_{}}} .
\end{eqnarray}
When $\gamma_w \gg \frac{m_\chi}{T}$, Eq.~\eqref{nchifilter} approaches $\frac{g_{\mathrm{DM}} T^3}{\pi^2}$ which is the equilibrium number density for Boltzmann distribution outside the bubble. This means that the bubble wall does not filter out DM particles at all in this limit. As $v_w \rightarrow 1$, $\gamma_w \gg 1$ and the exponent approaches $-\frac{m_\chi}{2\gamma_w T}$. On the contrary, in the case of $v_w \rightarrow 0$, $\gamma_w \rightarrow 1$ and the exponent becomes $-\frac{m_\chi}{T}$.
Then the penetration rate is
\begin{eqnarray}\label{Rin}
R_{\text {in }}^{}=\frac{1}{\gamma_w v_w}\left(\frac{{\gamma}_{w}\left(1-{v}_{w}\right) m_\chi^{} / T_{}+1}{4 {\gamma}_{w}^3\left(1-{v}_{w}\right)^2}\right) e^{-\frac{{\gamma}_{w}\left(1-{v}_{w}\right) m_\chi^{}}{T_{}}}\,\,.
\end{eqnarray}

\tc{Assessing the velocity of the bubble wall is crucial for understanding gravitational wave spectra, baryogenesis, and the dynamical mechanisms of dark matter~\cite{Moore:1995si,Jiang:2022btc}.} In the ultrarelativistic limit, the pressure on the bubble wall can be obtained from~\cite{Chway:2019kft,Bodeker:2009qy} :
\begin{equation}\label{friction}
P=\frac{d_n g_{\star} \pi^2}{90}\left(1+v_w\right)^3 \gamma_\omega^2 T^4 \,\,,
\end{equation}
which counts the difference of dof across the bubble wall. When the phase transition happens below the temperature of $0.1~\mathrm{GeV}$, then we choose $g_\star \approx 10$. And 
\begin{equation}
d_n \equiv \frac{1}{g_{\star}}\left[\sum_{0.2 m_i>\gamma_w T}\left(g_i^b+\frac{7}{8} g_i^f\right)\right],
\end{equation}
with $g_i^b$ and $g_i^f$ are the dof of the bosons and fermions respectively. If we only consider a complex scalar DM, we have $g_{\mathrm{DM}}=2$ and then $d_n \simeq 0.2$. As for super-cooling $\alpha \simeq \Delta U/\rho_{\mathrm{rad}}$, $v_w$ can be obtained by solving the balance condition, $\Delta U=P$ :
\begin{equation}\label{vwf}
\alpha=\frac{d_n}{3}\left(1+v_w\right)^3 \gamma_\omega^2\,\,.
\end{equation}
Note that this friction only works when $\gamma_w<0.2m_i/T$ such that it does not prohibit the run-away behavior of the bubble wall. The bubble wall velocity plays an essential role in the gravitational wave spectra. 

\begin{figure}[htbp]
	\centering
	\subfigure{
		\begin{minipage}[t]{0.5\linewidth}
			\centering
			\includegraphics[scale=0.5]{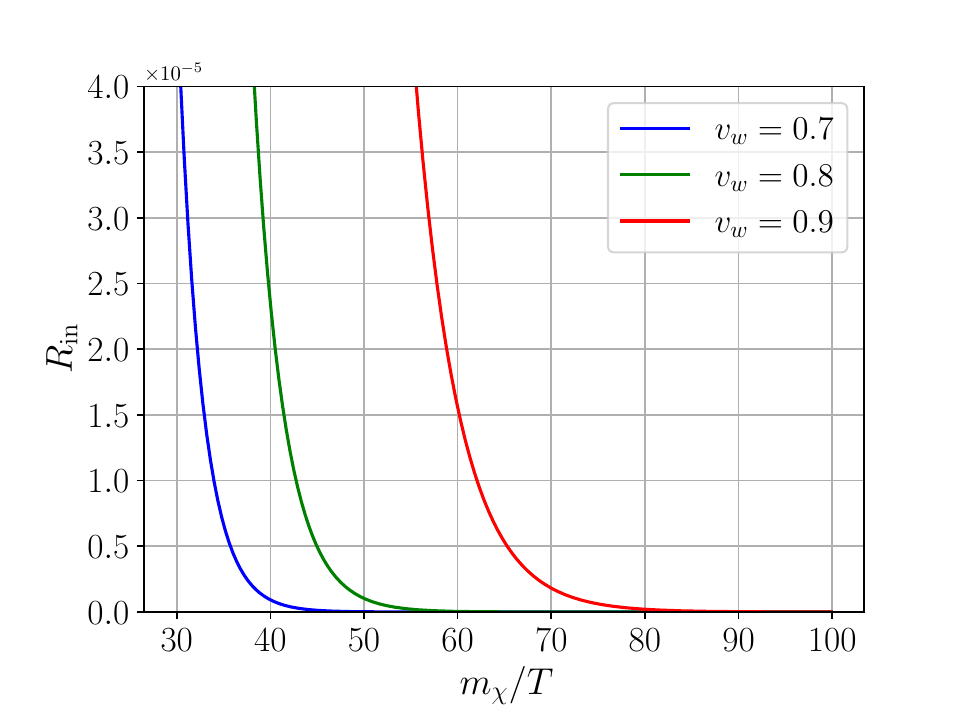}
	\end{minipage}}%
	\subfigure{
		\begin{minipage}[t]{0.5\linewidth}
			\centering
			\includegraphics[scale=0.5]{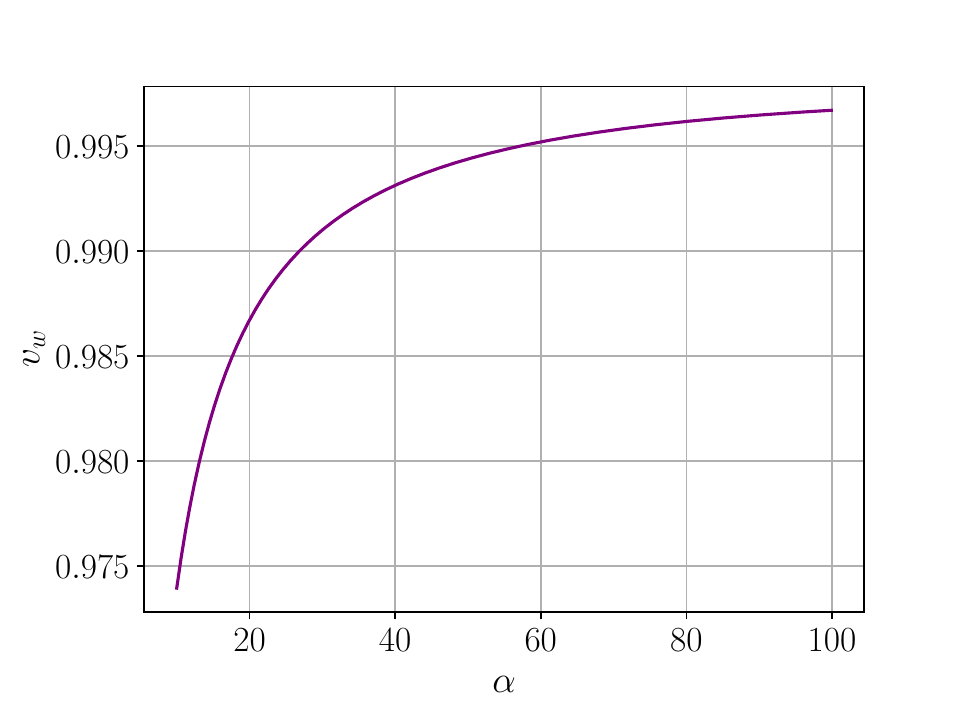}
	\end{minipage}}	
	\caption{Left: Penetration rate as function of $m_\chi/T$ for different bubble wall velocities. Right: Bubble wall velocity as a function of phase transition strength.}\label{vw}
\end{figure}
In the left panel of Fig.~\ref{vw}, the blue, green and red lines show the relation between penetration rate and $m_\chi/T$ with different bubble wall velocities. We can see that the penetration rate is positively correlated with velocity but negatively correlated with DM mass. Generally, we require $m_\chi/T \gtrsim 100$ to filter the DM effectively. The purple line in the right panel of Fig.~\ref{vw} shows the dependence of bubble wall velocity on phase transition strength $\alpha$, which is calculated by Eq.~\eqref{vwf}.

\section {THE dark matter candidates and their gravitational wave signals}\label{DC}
In this section, our primary focus lies on two distinct models of DM: Q-ball (Fermi-ball) DM and super-cool DM. These two models exhibit entirely different behaviors due to the variations in their penetration rates, as defined in Eq.~\eqref{Rin}. The Q-ball (Fermi-ball) DM or the filtered DM prefers the low penetration ability. The super-cool DM requires a large penetration rate instead.
\subsection{Q-ball and Fermi-ball dark matter}
\subsubsection{Relic density of Q-ball and Fermi-ball dark matter}
The so-called Q-ball is a non-topological compact soliton that exists in some models with a global symmetry and an associated conserved charge. To consider a simple model:
\begin{equation}\label{lag}
\mathcal{L}=\frac{1}{2}\left(\partial_\mu \phi\right)^2-U(\phi,T)+\left(\partial_\mu \chi\right)^*\left(\partial_\mu \chi\right)-y_\chi^2 \phi^2 \chi^* \chi\,\,,
\end{equation}
where $\chi$ is a complex scalar field. The particle $\chi$ gets its mass after it penetrates into a true vacuum after crossing the bubble wall,
\begin{equation}
    m_\chi=y_\chi v\,\,,
\end{equation}
where $v$ is the vacuum value deep inside the bubble.
These particles carry global charge, associated with the $U(1)$-symmetry $\chi \rightarrow \mathrm{e}^{i \theta} \chi$. The lowest energy state of a large enough charge is a spherical Q-ball with $\langle\phi\rangle=0$ inside and $\langle\phi\rangle=v$ outside. \tc{Here we assume that the charge asymmetry of $\chi$ and baryon asymmetry is related by $n_\chi = c_\chi n_B$, where $c_\chi$ is a constant at the time of Q-ball formation. }

We now discuss the phase transition process of Q-ball formation and then we get the number density and charge of Q-ball~\cite{Krylov:2013qe}. The lowest temperature at which remnants of the old phase can still form an infinite connected cluster is defined as $T_\star$. \tc{Referring to Eqs.~\eqref{Tp} and \eqref{IT}, it is observed that $p(T_\star) =  0.29$, indicating that the false vacuum fraction at $T_\star$ is 0.29. This corresponds to $I(T_\star) = 1.24$~\cite{Hong:2020est}.}

$T_\star$ is also the temperature when Q-balls start to form. Below the temperature $T_\star$, the old phase still constitutes approximately 0.29 of the universe's total volume, but it is subsequently divided into numerous ``false vacuum remnants". These remnants may further fragment into smaller pieces before ultimately shrinking into tiny Q-balls. The critical size, $r_\star$, of a remnant at the end of this fragmentation process is such that it shrinks to an insignificant size before another bubble containing the true vacuum can form within it~\cite{Krylov:2013qe}. It means~\cite{Hong:2020est}
\begin{equation}
	\Gamma\left(T_*\right) V_\star \Delta t \sim 1, \quad V_\star=\frac{4 \pi}{3} r_{\star}^3\,\,.
\end{equation}
where $\Delta t=r_{\star}/v_{w}$ is the time cost for shrinking. The number density of the remnants  $n_{Q}^{\star}$ can be expressed as: 
\begin{equation}
	n_{Q}^{\star}=\left(\frac{3}{4 \pi}\right)^{1 / 4}\left(\frac{\Gamma\left(T_\star\right)}{v_w}\right)^{3 / 4} p\left(T_{\star}\right)\,\,,
\end{equation}
since the condition $n_{Q}^\star V_\star=p(T_\star)$. In a remnant, the trapped Q-charge is
	$Q^{\star}=(1-R_{\mathrm{in}}) c_\chi \eta_B s_{\star}/{n_{Q}^{\star}}$
with $s_{\star}=2\pi^{2}g_{\star}T_{\star}^{3}/45$ and $\eta_{B}=0.9\times10^{-10}$~\cite{ParticleDataGroup:2018ovx} is the baryon asymmetry. Since $n_{Q} / s$ and $Q$ do not change in the adiabatic universe, at present they are
\begin{equation}
n_{Q}=\frac{n_{Q}^{\star}}{s_\star} s_0, \quad Q=Q^{\star}\,\,,
\end{equation}
with $s_0=2891.2~\mathrm{cm}^{-3}$ being the entropy density in current time~\cite{ParticleDataGroup:2018ovx}.

Q-ball is the
lowest energy state of large enough charge with $\phi=0$ inside and $\phi = v$
outside. For a Q-ball, its size $r_Q$ and energy $E$ are established by the equilibrium among the energy of $Q$ massless $\chi$-quanta confined in the potential well of radius $r_Q$, the surface energy term and the potential energy of the field $\phi$ in the interior~\cite{Krylov:2013qe}.
\begin{equation}
	E(r_{Q})=\frac{\pi Q}{r_{Q}}+4\pi\sigma_{0}r_{Q}^{2}+\frac{4 \pi}{3} r_{Q}^3 U_0\,\,,
\end{equation}
\tc{where $\sigma_{0}$ denotes the surface tension and $U_0\equiv \Delta U(T=0)$ is the difference of potential energy density between the false and true vacuum. The contribution of the surface term can be neglected in comparison to the volume term, primarily since these terms scale with $r_{Q}^{2}$ and $r_{Q}^{3}$, respectively.} And the Q-ball parameters $r_{Q}$ and $m_Q$ are obtained by minimizing its energy above:
\begin{equation}
   r_Q=\left(\frac{Q}{4 U_0}\right)^{1 / 4}, \quad m_Q=\frac{4 \sqrt{2} \pi}{3} Q^{3 / 4} U_0^{1 / 4}\,\,.
\end{equation}
 The Q-ball is stable provided that its energy is smaller than the rest energy of $Q$ massive $\chi$-quanta in the vacuum $\phi=v$,
\begin{equation}
	m_Q<m_\chi Q\,\,.
\end{equation}

Given an effective potential $U(\phi,T)$, it is convenient to rewrite Q-ball profile in terms of the action at $T_\star$. Due to $\Gamma(T_{\star}) / T_{\star}^4 \approx e^{-S_3\left(T_{\star}\right) / T_{\star}}$, the mass, radius and charge of a Q-ball can be expressed as 
\begin{equation}
	\begin{aligned}
		m_{Q} \approx  9.8\times10^{-6} \times {v_w}^{9 / 16}\times U_0^{1 / 4}\times\left[(1-R_{\mathrm{in}})c_\chi\right]^{3 / 4}\times \exp \left\{\frac{9}{16}\left(\frac{S_3\left(T_{\star}\right)}{T_{\star}}\right)\right\}, 
	\end{aligned}
\end{equation}
\begin{equation}
	\begin{aligned}
		r_{Q} \approx  8.4\times10^{-3} \times {v_w}^{3 / 16}\times U_0^{-1 / 4}\times\left[(1-R_{\mathrm{in}})c_\chi\right]^{1 / 4}\times \exp \left\{\frac{3}{16}\left(\frac{S_3\left(T_{\star}\right)}{T_\star}\right)\right\}, 
	\end{aligned}
\end{equation}
\begin{equation}
	\begin{aligned}
		Q \approx  1.95\times10^{-8} \times {v_w}^{3 / 4}\times\left[(1-R_{\mathrm{in}})c_\chi\right]\times \exp \left\{\frac{3}{4}\left(\frac{S_3\left(T_\star\right)}{T_\star}\right)\right\}. 
	\end{aligned}
\end{equation}
The number density of Q-balls is
\begin{equation}
	\begin{aligned}
		n_Q \approx 4.6\times10^{-3}\times s_0 \times {v_w}^{-3 / 4}\times \exp \left\{-\frac{3}{4}\left(\frac{S_3\left(T_\star\right)}{T_\star}\right)\right\}. 
	\end{aligned}
\end{equation}
The relic density of Q-balls is
\begin{equation}
	\Omega_{Q} h^2=\frac{n_{Q} m_{Q}}{\rho_{\mathrm{c}}} h^2,
\end{equation} 
The critical energy density, denoted as $\rho_{\mathrm{c}}$, is given by $3 H_{0}^{2} M_{\mathrm{Pl}}^2 /(8 \pi)$, which equals $(2.5 \times 10^{-3}~\mathrm{eV})^{4}$. Here, $H_0$ is the Hubble constant today, and $h=H_{0} /\left(100~\mathrm{km} \cdot \mathrm{s}^{-1} \cdot\mathrm{Mpc}^{-1}\right)\approx 0.67$. Then we can get
\begin{equation}
	\begin{aligned}
		\Omega_{Q}h^2 \approx  11.305 \times {v_w}^{-3 / 16}\times\left(\frac{U_0^{1 / 4}}{1~\mathrm{GeV}}\right)\times\left[(1-R_{\mathrm{in}})c_\chi\right]^{3 / 4}\times \exp \left\{-\frac{3}{16}\left(\frac{S_3\left(T_\star\right)}{T_\star}\right)\right\}\,\,.
	\end{aligned}
\end{equation}
Because we generally impose $m_\chi/T_\star \gtrsim 25$ in order to trap some DM particles into the false vacuum, the DM particles that penetrate into the bubble are out-of-equilibrium. Then the DM particles won't experience the freeze-out process. However, if the escaping part is dominant, the DM relic abundance is given by the excess of $\chi$ over $\bar{\chi}$,
\begin{equation}
\Omega_\chi h^2  =R_{\mathrm{in}} c_\chi \eta_B s_0 m_\chi  =0.0036 \times\left(\frac{R_{\mathrm{in}}}{0.1}\right)\left(\frac{c_\chi}{1.46}\right)\left(\frac{m_\chi}{1 ~\mathrm{GeV}}\right)
\end{equation}
and $\Omega_{\overline{\chi}} h^2=0$.
In our model, $U_0 \simeq 43.5 M^4$. As an example, we find that the value of $c_\chi$ is approximately \tc{between $10^{14}$ and $10^{15}$ to achieve the observed DM density of $\Omega_{\mathrm{Q}} h^2 \simeq 0.12$ when $v_w=0.9$ and $M= 10~\mathrm{MeV}$ and we find $\Omega_{\mathrm{Q}} h^2 < \Omega_\chi h^2$ at this point}. \tc{However, if the bubble wall velocity is lower, the Q-ball can still provide the dominant contribution of DM. }

However, if the DM particle $\chi$ is fermion, it can form Fermi-ball DM when it carries a conserved global charge. In the Fermi-ball model, the DM relic density is~\cite{Hong:2020est}:
\begin{equation}
	\Omega_{\mathrm{FB}}h^2=0.12 \times (1-R_{\mathrm{in}})\left(\frac{c_\chi}{1.46}\right)\left(\frac{U_0^{1 / 4}}{1~\mathrm{GeV}}\right)\,\,,
\end{equation}
where we choose $S_3(T_\star)/T_\star=140$. For example, to meet the DM relic density, a value of approximately $c_\chi \simeq 146$ is required for $M= 10~\mathrm{MeV}$, which appears to be a more natural requirement. And in this case, $\Omega_\chi h^2 \ll \Omega_{\mathrm{FB}}h^2$ is satisfied. The main difference between the Q-ball and Fermi-ball is that for a Q-ball, the mass is proportional to $Q^{3/4}$, whereas for a Fermi-ball, the mass is directly proportional to $Q$.

\begin{figure}[htbp]
	\begin{center}
		\includegraphics[width=0.8\linewidth]{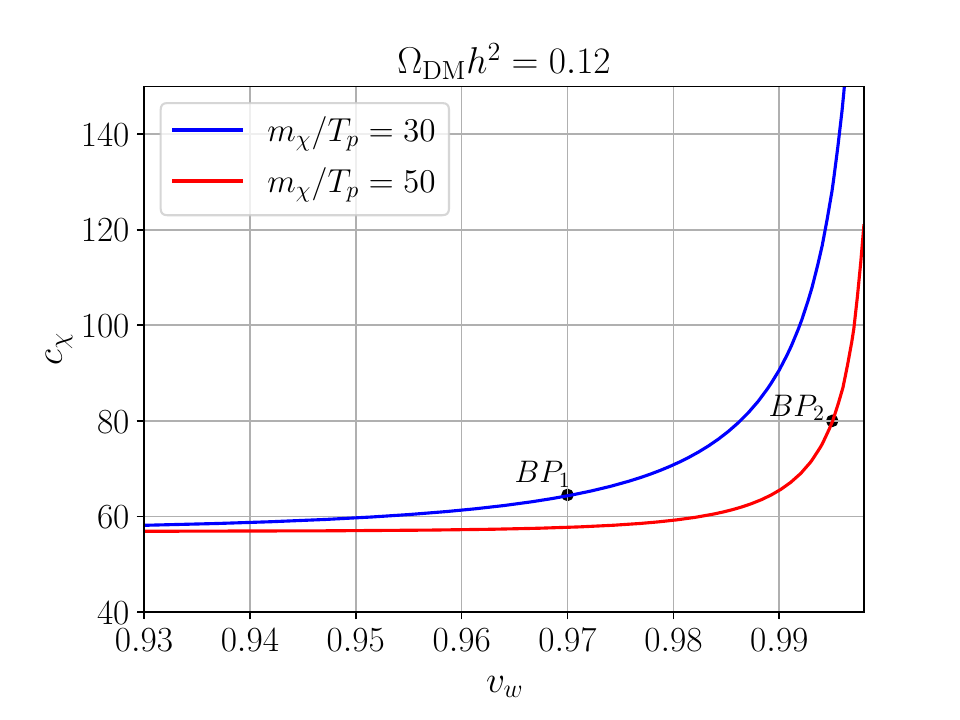}
		\caption{Contour plot of $c_\chi$ and $v_w$ of Fermi-ball DM for $\Omega_{\mathrm{DM}}h^2=0.12$. Here we choose $M=10~\mathrm{MeV}$ and $S_3\left(T_\star\right)/T_\star=140$. We take the approximation $T_\star \approx T_p$.}
		\label{QB}
	\end{center}
\end{figure}

In Fig.~\ref{QB} we show the contour plot of $v_w$ and $c_\chi$ for Fermi-ball DM which agrees with the observed relic density $\Omega_{\mathrm{DM}}h^2=0.12$. \tc{We choose $M=10~\mathrm{MeV}$. This choice is guided by the need to balance two critical aspects: firstly, if the mass scale $M\equiv e^{-\frac{1}{3}+\gamma_E} \Lambda_u / 4 \pi$ is significantly less than 10 MeV, it would result in a phase-transition temperature lower than 4 MeV, potentially destroying BBN. Secondly, if $M$ substantially exceeds 10 MeV, we found that it is hard to form Q-ball or Fermi-ball during a FOPT. Furthermore, a much larger $M$ will lead to a high phase-transition strength $(\alpha\gg1)$, which could result in large super-cooling, thereby diluting the DM relic density. This is the main point of super-cool DM scenario which we will discuss in the next subsection.} We choose two cases: $m_\chi/T_p=30$ and $m_\chi/T_p=50$ respectively. Here we take the approximation $T_\star \simeq T_p$. It can be seen that for smaller $m_\chi/T_{p}$, the Fermi-ball DM is more difficult to form at large bubble wall velocity since the smaller mass indicates a weaker filtering ability. In other words, DM particles with smaller masses are more challenging to confine within the bubbles.

If the filtering ability is weaker, we expect that more DM particles can penetrate into the bubbles. This can produce so-called filtered DM which can evade the unitary bound of the freeze-out mechanism~\cite{Baker:2019ndr}. In our work, we do not take it into account since we find large $m_\chi/T_p \gtrsim 100$ often indicates a huge phase-transition strength $\alpha_p$ then a huge bubble wall velocity $v_w \simeq 1$. All DM particles will penetrate into the bubble in this case. However, large super-cooling inevitably leads to inflation in the phase transition process which can produce super-cool DM.

\subsubsection{Gravitational wave spectra of Fermi-ball dark matter}
\begin{figure}[htbp]
	\begin{center}
		\includegraphics[width=0.8\linewidth]{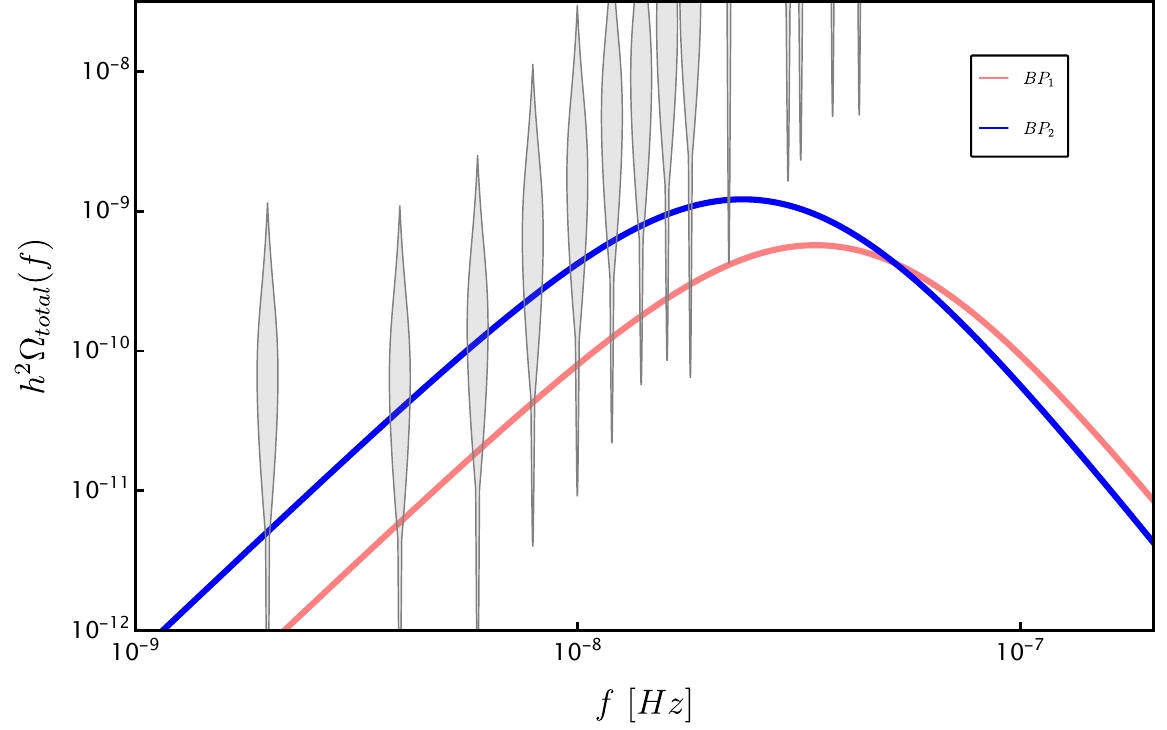}
		\caption{Gravitational wave spectra for two benchmark parameters of Fermi-ball DM. The grey violins shaded region represents the observation of the NANOGrav~\cite{NANOGrav:2023hvm} of nHz 
        gravitational waves.}
		\label{B}
	\end{center}
\end{figure}

The SGWB is given by three different contributions: bubble collision, sound wave and turbulence. The details of the gravitational wave spectra are shown in Appendix.~\ref{GWspe}. We can see that the peak frequency is proportional to $\beta/H_p \times T_{\mathrm{RH}}$ and is inversely proportional to bubble wall velocity $v_w$. To avoid the BBN constraints the reheating temperature should be larger than $4~\mathrm{MeV}$. This temperature depends on the decay width of the scalar field such that we can set it as a free parameter $T_{\mathrm{RH}}=5~\mathrm{MeV}$. The details of the reheating temperature will be discussed in the next subsection. In order to fit the spectra of SGWB, one can only decrease the $\beta/H_p$ and increase $v_w$. These can be naturally realized in super-cooling phase transition. 
We choose two benchmark points that fulfill $\Omega_{\mathrm{DM}}h^2=0.12$ for $m_\chi/T_p=30$ and $m_\chi/T_p=50$ respectively as shown in TABLE.~\ref{ptableQ}. Due to the asymptotic behavior of friction at high bubble wall velocity, $P \sim \gamma_w^2$ in Eq.~\eqref{friction}, we find the bubble wall hardly run away. Then the main contribution to SGWB comes from the sound wave. The gravitational wave spectra are given in Fig.~\ref{B}. As we can see in Fig.~\ref{B}, the SGWB of the Fermi-ball DM can fit the data of NANOGrav at the low-frequency range.

\begin{table}[t]
	\centering
	
	\setlength{\tabcolsep}{3mm}
	
	\begin{tabular}{c|c|c|c|c|c|c|c|c|c}
		\hline\hline
		& $y_A$ & $M$~[MeV] &$T_{p}$~[MeV] &$v_p/T_p$ &$\alpha_p$ & $\beta/H_p$ & $y_\chi$ & $c_\chi$ & $v_w$\\
		\hline
        $BP_1$	 &0.88 & 10 & 7  &12 &8.4 & 79.7 & 2.5 & 67 &0.97\\
            \hline
		$BP_2$	 &0.78 & 10 & 11 &21 & 50 & 74.1 & 2.4 & 81 & 0.995\\
		\hline\hline
	\end{tabular}
	\caption{Two sets of benchmark parameters that fulfill $\Omega_{\mathrm{DM}} h^2=0.12$ and the corresponding phase transition parameters in Fermi-ball DM. The value of $y_\chi$ is to fulfill $m_\chi/T_p = 30, 50$ in $BP_1$ and $BP_2$ respectively. }\label{ptableQ}
\end{table}

\subsection{Super-cool dark matter}\label{SC}
\subsubsection{Relic density of super-cool dark matter}
On the contrary, we will discuss another DM, called the super-cool DM mechanism~\cite{Hambye:2018qjv} in which a significant amount of DM particles can penetrate the bubbles. The gauge boson $A$ in our model is the natural DM candidate. The number density of particles is diluted during the inflation induced by a strong super-cooling phase transition. Unlike the standard freeze-out mechanism, the DM particles are in thermal equilibrium with the SM plasma only in the false vacuum. The DM particles penetrating the bubble depart from thermal equilibrium instantly as they gain huge mass across the bubble wall. The SM particles lighter than the reheating temperature can easily thermalize, however, DM particles can not. This mechanism can naturally suppress the DM density to satisfy the observed DM relic density.

The scale-invariant model may produce strong super-cooling. If this happens, the vacuum energy may dominate the universe. Then the universe will experience inflation. The thermal inflation begins at temperature $T_{\rm infl}$ and induces a constant expansion rate,
\begin{eqnarray}\label{Tinfl}
	\frac{g_{\star}\pi^2 T_{\rm infl}^4}{30} = \Delta U(T_{\rm infl}) =\frac{3H_{\rm infl}^2M_{\mathrm{pl}}^2}{8\pi}\,\,.
\end{eqnarray}
The scale factor during inflation will grow as $a(t)=a_{\rm infl}e^{H_{\rm infl}t}$, and the temperature drops as $T = T_{\rm infl} a_{\rm infl}/a$ due to the entropy conservation.

The inflation in some classical scale-invariant models can be ceased by the QCD phase transition at $T_{\mathrm{QCD}}\sim 100~\mathrm{MeV}$. The quark condensate induces a linear term in the Higgs potential such that the Higgs acquires a $T$-dependent vacuum expectation value. Then if the phase transition field in the model has a nonzero mixing with the Higgs $-\lambda_{\Phi H}|\Phi|^2 |H|^2$, this will also induce the negative mass term of $\phi$ which triggers the end of the inflation. In this work, we choose benchmark points that the inflation ends at $T_p$.

After the phase transition, the vacuum energy stored in the scalar field will be transferred to particles. The ultimate temperature of this reheating process is determined by the energy transfer rate $\Gamma_{\phi}$ (see Appendix \ref{RH}),
\begin{eqnarray}\label{TRH}
	T_{\rm RH} = T_{\rm infl}\times \text{min} (1,\frac{\Gamma_\phi}{H_{\rm infl}})^{1/2}\,\,.
\end{eqnarray}
The DM number density can be traced from the thermal history,
\begin{eqnarray}
	n_{A}(T_{\rm RH}) \simeq \left(\frac{a_{p}}{a_{\rm RH}}\right)^3 n_{A}^{\rm true}(T_{p}) \,\,,
\end{eqnarray}
where $n_{A}^{\rm true}(T_{p})$ is the DM number density that just enters the bubble wall. Note that before and after the reheating, we have energy conservation,
\begin{eqnarray}
	\Delta U(T_{p})\times a_{p}^3 = \frac{g_{\star}\pi^2 T_{\rm RH}^4}{30} \times a_{\rm RH}^3\,\,.
\end{eqnarray}
Use $\Delta U(T_{p})\simeq\Delta U(T_{\rm infl})$, then we have
\begin{eqnarray}\label{nchi}
	n_{A}(T_{\rm RH}) =  \frac{T_{\rm RH}^4}{T_{\rm infl}^4} \times n_{A}^{\rm true}(T_{p}) =  \frac{T_{\rm RH}^4}{T_{\rm infl}^4}\times \frac{T_{p}^3}{T_{\rm infl}^3} \times n_{A}^{\rm false}(T_{\rm infl})\,\,,
\end{eqnarray} 
where in the second equality we have used the fact that all DM transfer into the bubble wall and $a_{\rm infl}/a_{p}=T_{p}/T_{\rm infl}$. We then get
\begin{eqnarray}
	Y_{A}^{\rm RH} = Y_{A}^{\rm eq}\frac{T_{\rm RH}}{T_{\rm infl}}\left(\frac{T_p}{T_{\rm infl}}\right)^3 , \quad Y_{A}^{\rm eq}=\frac{n_{A}^{\rm false}(T_{\rm infl})}{s(T_{\rm infl})}=\frac{45g_{A}}{2\pi^4g_{\star}}\,\,,
\end{eqnarray}
which is consistent with the result of \cite{Hambye:2018qjv}. Here $Y_A= n_A/s$ and  $g_{\mathrm{A}}=3$ which is the dof of DM.

One may worry about that if the scalar field $\phi$ can reheat the SM plasma. This can not be done by the Higgs portal as we have forbidden the mixing of terms between them. However, the reheating can be realized by the direct coupling with electrons and photons~\cite{Madge:2023cak}. If the reheating temperature can exceed the decoupling temperature of the DM $T_{\mathrm{RH}} \gtrsim T_{\mathrm{dec}}$. The DM can be thermalized again and the supercooled population will be erased. Then the DM relic density will be set by the standard freeze-out.  If instead $T_{\mathrm{RH}} \lesssim T_{\mathrm{dec}}$ the supercooled population remains basically unchanged but the second population is produced by sub-thermal process~\cite{Hambye:2018qjv}. The evolution of the sub-thermal population follows the Boltzmann equation,
\begin{equation}
\frac{d Y_{A}}{d x}=-\frac{\lambda_a}{x^2}\left(Y_{A}^2-(Y_{A}^{\mathrm{eq}})^2\right), \quad \text { with } \quad x=\frac{m_{A}}{T} .
\end{equation}
where $\lambda_a=M_{\mathrm{Pl}} m_{A}\left\langle\sigma_{\mathrm{ann}} v_{\mathrm{rel}}\right\rangle \sqrt{\pi g_{\star} / 45}$. Here we choose $\left\langle\sigma_{\mathrm{ann}} v_{\mathrm{rel}}\right\rangle =\left\langle\sigma_{\mathrm{ann}} v_{\mathrm{rel}}\right\rangle_{AA\leftrightarrow \phi \phi } = \frac{11y_A^4}{6912\pi m_A^2}$ which is the thermal-averaged $s$-wave
cross-section for annihilation of DM particle $A$~\cite{Hambye:2018qjv}. For $T_{\mathrm{RH}} \lesssim T_{\mathrm{dec}}$ the initial super-cooled population of DM can be ignored, then the sub-thermal contribution is
\begin{equation}
\left.Y_{A}\right|_{\text {sub-thermal }}=\lambda_a \int_{x_{\mathrm{RH}}}^{\infty} \frac{d x}{x^2} (Y_{A}^{\mathrm{eq}})^2=\lambda_a \frac{2025 g_{A}^2}{128 \pi^7 g_{\star}^2} e^{-2 x_{\mathrm{RH}}}\left(1+2 x_{\mathrm{RH}}\right)\,\,.
\end{equation}
where we have used $Y_{A}^{\mathrm{eq}}(x)\equiv \frac{45}{2^{5/2}\pi^{7/2}}\frac{g_A}{g_\star} x^{3/2}e^{-x}$ in the non-relativistic limit.

%

Note that if we release the assumption that all DM can pass into the bubble in Eq.~\eqref{nchi}, the population of super-cool DM should be
\begin{eqnarray}
	\left.Y_{A}\right|_{\text {super-cool }} \equiv Y_{A}^{\rm RH} = Y_{A}^{\rm eq}\frac{T_{\rm RH}}{T_{\rm infl}}\left(\frac{T_{p}}{T_{\rm infl}}\right)^3 \times R_{\rm in}\,\,,
\end{eqnarray}
where $R_{\rm in}$ comes from Eq.~\eqref{Rin} which quantifies the filtering effect at $T_p$. As we already see, the filtering effects mainly come from the competition between particle typical energy $\gamma_w T_{p}$ and the particle mass $m_A^{}$. On one hand, strong super-cooling will give a large $m_A^{}/T_{p}$. However,  it may also lead to a large bubble wall velocity. In this work, we found for $\alpha_p \gg 100$, the bubble wall velocity is very close to 1. So we set $v_w=1$ in the subsequent calculation such that $R_{\mathrm{in}}=1$.

The current relic density consists of both the super-cooled population and the sub-thermal population
\begin{equation}
\Omega_{\mathrm{DM}}h^2 = \frac{(\left.Y_{A}\right|_{\text {super-cool }}+\left.Y_{A}\right|_{\text {sub-thermal }})m_As_0}{\rho_c}h^2\,\,.
\end{equation}

The reheating temperature $T_{\mathrm{RH}}$ is also used in SGWB spectra. If the reheating is instant, the $T_{\mathrm{RH}}$ can be expressed as
\begin{equation}
    T_{\mathrm{RH}} = \left(\frac{30\Delta U(T_{\mathrm{infl}})}{g_\star\pi^2}\right)^{1/4}= \left(\frac{30\Delta U(T_{p})}{g_\star\pi^2}\right)^{1/4} \simeq (1+\alpha_p)^{1/4}T_p\,\,,
\end{equation}
where we used the fact that $\alpha_p \gg 1$ for super-cooling phase transition.
However, if we consider the finite decay width of the scalar field as considered in Eq.~\eqref{TRH}, the reheating temperature will also be suppressed also the peak frequency of the gravitational wave spectra. 
\begin{equation}
T_{\mathrm{RH}}=\left(\frac{90}{8\pi}\right)^{1/4} g_\star^{-1 / 4} \sqrt{M_{\mathrm{pl}} \Gamma_\phi}
\end{equation}
This can help us to relate the low-frequency gravitational wave spectra of $T_\mathrm{RH} \simeq \mathcal{O}(10)$~MeV with the high-energy physics at $\mathcal{O}(100)$~GeV. We choose $T_{\mathrm{RH}}=20,15$~MeV which correspond to $\Gamma_\phi =5.5\times 10^{-20},3.1\times 10^{-20} $~MeV respectively. Note that in order to avoid the BBN constraint we should have $\Gamma_\phi \gtrsim 6.9\times 10^{-21}$~MeV so that $T_{\mathrm{RH}} \gtrsim 4$~MeV. The decoupling temperature of DM is generally $T_{\mathrm{dec}} \simeq m_{\mathrm{DM}}/25$. The $BP_{1-4}$  correspond to $m_A/T_{p} \gg 25$ where the thermal population can be ignored and super-cool population dominates.

\subsubsection{Gravitational wave spectra of super-cool dark matter}\label{GWs}

\begin{figure}[htbp]
	\begin{center}
		\includegraphics[width=0.8\linewidth]{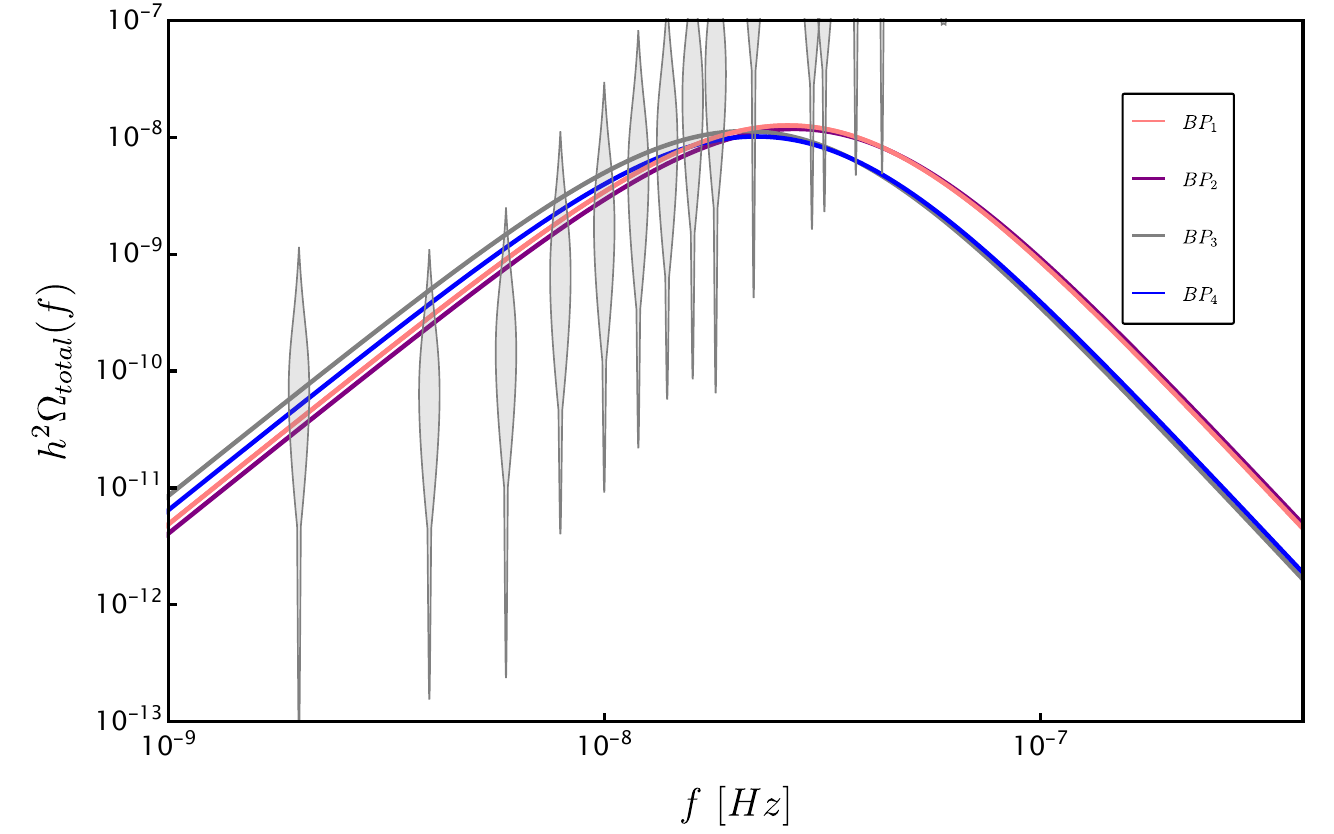}
		\caption{Gravitational wave spectra for four benchmark parameters of super-cool DM. The grey violins shaded region represents the observation of the NANOGrav of nHz gravitational waves.}
		\label{A}
	\end{center}
\end{figure}

\tc{We choose four benchmark points for super-cool DM which fulfills DM relic density $\Omega_{\mathrm{DM}}h^2=0.12$ as shown in TABLE.~\ref{ptable}. We also show corresponding phase-transition parameters.} The judgment of run-away condition refers to \cite{Levi:2022bzt,Azatov:2019png}. We find the bubble wall hardly run away at these points. The main contribution to SGWB comes from the sound wave. In Fig.~\ref{A}, we show that the gravitational wave spectra for these benchmark parameters can fit the PTA observation well. The grey violins shaded region represents the distribution of the NANOGrav signals~\cite{NANOGrav:2023hvm}. The pink, purple, grey and blue dashed lines show the total gravitational wave power spectra for these benchmark parameters. As we can see, the super-cooling phase transition can be detected by PTA experiments without violating the BBN bound. The super-cooling phase transition is a natural solution to avoid the BBN constraints. 

\begin{table}[t]
	\centering
	
	\setlength{\tabcolsep}{3mm}
	
	\begin{tabular}{c|c|c|c|c|c|c|c|c}
		\hline\hline
		& $y_A$ & $M$~[GeV] &$T_{\mathrm{RH}}$~[MeV] &$T_p$~[GeV] &$\alpha_p$ & $\beta/H_p$ \\
		\hline
        $BP_1$	 &0.622 & 380 & 20 &4.02  &$1.06 \times 10^8$ & 15.4 \\
            \hline
		$BP_2$	 &0.62 & 270 & 20 &2.83  &$1.09 \times 10^8$ & 16.0 \\
            \hline
		$BP_3$	&0.621 & 240 & 15 & 2.71 &$8.11\times10^7$  & 16.4 \\
            \hline
		$BP_4$	&0.62 & 110 & 15 & 1.45  &$4.37 \times 10^7$ & 17.4 \\
		\hline\hline
	\end{tabular}
	\caption{Four sets of benchmark parameters which fulfill $\Omega_{\mathrm{DM}}h^2=0.12$ and the corresponding phase transition parameters in super-cool DM mechanism. }\label{ptable}
\end{table}

\section{Conclusion}\label{con}
We have studied the implication of the possible observation of SGWB in
the frame of dynamical DM formed through a \rc{dark} FOPT in the early universe. After investigating several representative dynamical DM formation mechanisms, we find that a  \rc{dark} FOPT with strong super cooling can fit the PTA observation, which might be a hint for a generic class of dynamical DM mechanisms. The super-cooling phase transition has two essential characteristics: large $v_p/T_p$ and large $T_c/T_p$. The former induces a strong filtering effect and the latter indicates low-scale inflation. As the DM mass is huge during a strong super-cooling phase transition, the DM particles are trapped in the false vacuum and form the Q-ball DM or Fermi-ball DM. As the inflation happens, the DM relic density can be naturally suppressed which is so-called super-cool DM. These two dynamical DM models are differentiated by penetration rates. \tc{The penetration rate is characterized by the vacuum expectation value over temperature $v_p/T_p$ and bubble wall velocity $v_w$. Both of them are model-dependent. The Q-ball or Fermi-ball scenario prefers large $v_p/T_p$ and small $v_w$ which leads to ultra-small penetration rate $R_{\mathrm{in}}\simeq 0$. Then the Q-ball or Fermi-ball, as the false vacuum remnants, could be the DM candidate. As the super-cooling is strong enough $\alpha \gg 1$, we have $v_w \rightarrow 1$ and the corresponding penetration rate $R_{\mathrm{in}} \rightarrow 1$. In this case, although almost all DM particles penetrate into the bubbles, the DM number density can be naturally diluted during the inflation epoch of super-cooling phase transition.}

\tc{For Q-ball DM with a low penetration rate, we have found that in order to fulfill the DM relic density, the charge asymmetry of DM particles has to be several orders of magnitude larger than the baryon asymmetry. In contrast, Fermi-ball DM, whose charge asymmetry is comparable with the baryon asymmetry, is a more natural DM candidate.} In the case of a strong super-cooling phase transition where DM particles have a high penetration rate, we discuss the super-cool DM mechanism. Particularly, we found for super-cool DM mechanism, the DM relic density and the SGWB spectra can be fitted simultaneously if we have an appropriate reheating temperature $T_{\mathrm{RH}}$ or the corresponding finite decay with scalar field $\Gamma_\phi$ which is model-dependent. The reheating temperature of super-cool DM mechanism in a specific model should be discussed in detail. \tc{Notably, super-cool DM fits the SGWB spectra better, owing to its larger phase-transition strength $\alpha$ and smaller phase-transition rate $\beta / H$. The $ \alpha $ and $ \beta/H $ determine the strength and the peak frequency of gravitational wave spectra, respectively.} \tc{In this work, we discuss several dynamical DM mechanisms which are differentiated by their different penetration rates. More discussions on how to distinguish different dynamical DM models and their implications on the nature of DM are left in our future study.}

\begin{acknowledgments}
This work was supported by the National Natural Science Foundation of China (NNSFC) under Grant No. 12205387 and by Guangdong Major Project of Basic and Applied Basic Research (Grant No. 2019B030302001).
\end{acknowledgments}

\appendix

\section{Reheating temperature}\label{RH}
As the inflaton experiences a period of coherent oscillation before decaying to particles, the Hubble rate in this period has the form,
\begin{eqnarray}
	H(a) = \sqrt{\frac{8\pi}{3}\frac{\rho_{\rm infl}}{M_{\rm pl}^2}\left(\frac{a_{\rm ini}}{a}\right)^3} \,\,,
\end{eqnarray}
where $\rho_{\rm infl}$ is the energy density which is stored in inflaton, i.e., the vacuum energy density. $a_{\rm ini}$ is the scale factor at the beginning of coherent oscillation, in our case, the end of the phase transition, $a_{\rm ini}=a_{p}$. The reheating process will begin when the decay rate equals the Hubble rate,
\begin{eqnarray}
	\Gamma_{\phi} = H(a_{\rm dec})\,\,.
\end{eqnarray}
Then the energy stored in inflaton at decaying is
\begin{eqnarray}
	\rho_{\rm dec}=\frac{3M_{\rm pl}^2 \Gamma_{\phi}^2}{8\pi}\,\,.
\end{eqnarray}
At the end of reheating, the energy will be entirely transferred to the radiation, and then we have,
\begin{eqnarray}
	\rho_{\rm dec}=\rho_{\rm RH}=\frac{\pi^2 g_\star}{30}T_{\rm RH}^4\,\,.
\end{eqnarray}	
As a consequence,
\begin{eqnarray}
	T_{\rm RH} = \left(\frac{45}{4\pi^3 g_\star}\right)^{1/4}\sqrt{\Gamma_{\phi}M_{\rm pl}}\,\,.
\end{eqnarray}

If the reheating is instant, we will have
\begin{eqnarray}
	\rho_{\rm dec} = \rho_{\rm infl} = \frac{3 M_{\rm pl}^2}{8\pi} H_{\rm infl}^2 =\frac{\pi^2 g_\star}{30}T_{\rm RH}^4\,\,,
\end{eqnarray}
where $H_{\rm infl}=H(a=a_{\rm infl})$ is the Hubble rate before inflaton oscillation. Then 
\begin{eqnarray}
	T_{\rm RH} \equiv T_{\rm RH}^{\mathrm{max}} = \left(\frac{45}{4\pi^3 g_\star}\right)^{1/4}\sqrt{H_{\rm infl}M_{\rm pl}}\,\,.
\end{eqnarray}

In summary, we have the expression of reheating temperature,
\begin{eqnarray}
	T_{\rm RH} = T_{\rm infl}\times\text{min} (1,\frac{\Gamma_\phi}{H_{\rm infl}})^{1/2}\,\,.
\end{eqnarray}
with $T_{\rm infl}$ defined in Eq.~\eqref{Tinfl}.

\section{Gravitational waves spectrum}\label{GWspe}
The SGWB of a FOPT from three mechanisms (bubble collision, sound wave and turbulence) can be obtained using the dynamic parameters of the FOPT.
\begin{itemize}
	\item Bubble collision\\
	The numerical simulation provides the formula of the SGWB spectrum from bubble collisions~\cite{Huber:2008hg,Caprini:2015zlo} at the reheating temperature $T_{\rm RH}$~, where we account for the impact of redshift:
	\begin{equation}
		h^2 \Omega_{\mathrm{co}}(f) \simeq 1.67 \times 10^{-5}\left(\frac{H_p}{\beta}\right)^2\left(\frac{\kappa_\phi \alpha_p}{1+\alpha_p}\right)^2\left(\frac{100}{g_{\star}}\right)^{1 / 3} \frac{0.11 v_w^{3}}{0.42+v_w^2} \frac{3.8\left(f / f_{\mathrm{co}}\right)^{2.8}}{1+2.8\left(f / f_{\mathrm{co}}\right)^{3.8}}\,\,,
	\end{equation}
	where $v_{w}$ is bubble wall velocity, $H_p$ denotes the Hubble rate at percolation temperature, while $ \kappa_\phi $ represents the percentage of vacuum energy converted into the scalar fields' gradient energy. The peak frequency of bubble collision processes is :
	\begin{equation}
		f_{\mathrm{co}} \simeq 1.65 \times 10^{-5} \mathrm{~Hz} \left(\frac{\beta}{H_p }\right)\left(\frac{0.62}{1.8-0.1 v_w+v_w^2}\right)\left(\frac{T_{\rm RH}}{100~ \mathrm{GeV}}\right)\left(\frac{g_\star}{100}\right)^{1 / 6}\,\,.
	\end{equation}
	\item Sound wave\\
	The sound waves can produce more significant and durable SGWB. The formula of the SGWB spectrum from sound waves is~\cite{Hindmarsh:2017gnf}:
	\begin{equation}
		\begin{aligned}
			h^2 \Omega_{\mathrm{sw}}(f) \simeq & 2.65 \times 10^{-6} \Upsilon_{\mathrm{sw}} \left(\frac{H_p}{\beta}\right)\left(\frac{\kappa_v \alpha_p}{1+\alpha_p}\right)^2\left(\frac{100}{g_{\star}}\right)^{1 / 3} 
			  v_w  \left(f / f_{\mathrm{sw}}\right)^3  \left(\frac{7}{4+3\left(f / f_{\mathrm{sw}}\right)^2}\right)^{7 / 2}\,\,,
		\end{aligned}
	\end{equation}
where $  \kappa_v $ reflects the percentage of vacuum energy that results in the fluid's bulk motion. And the peak frequency of sound waves processes is  :
\begin{equation}
		f_{\mathrm{sw}} \simeq 1.9 \times 10^{-5} \mathrm{~Hz} \frac{1}{v_w}\left(\frac{\beta}{H_p}\right)\left(\frac{T_{\rm RH}}{100~ \mathrm{GeV}}\right)\left(\frac{g_\star}{100}\right)^{1 / 6}\,\,.
\end{equation}
We have included the suppression factor of the short period of the sound wave,
\begin{equation}
\Upsilon_{\mathrm{sw}} = \left(1-\frac{1}{\sqrt{1+2\tau_{\mathrm{sw}}H_p}}\right)\,\,,
\end{equation}
where
\begin{equation}
    \tau_{\mathrm{sw}}H_p \approx (8\pi)^{\frac{1}{3}}\frac{v_w H_p}{\beta \sqrt{3\kappa_v\alpha/(4+4\alpha)}}\,\,.
\end{equation}
\item Turbulence  \\
The formula of the SGWB spectrum from turbulence is~\cite{Binetruy:2012ze}:
\begin{equation}
		h^2 \Omega_{\mathrm{turb}}(f) \simeq 3.35 \times 10^{-4} \left(\frac{H_p v_w}{\beta}\right)\left(\frac{\kappa_{\mathrm{turb}} \alpha_p}{1+\alpha_p}\right)^{3 / 2}\left(\frac{100}{g_\star}\right)^{1 / 3} \frac{\left (f / f_{\text {turb }}\right)^3}{\left(1+f / f_{\text {turb }}\right)^{11 / 3}\left(1+8 \pi f / h_p\right)}\,\,,
\end{equation}	
and $h_p$ is expressed as:
 \begin{equation}
h_p=1.65 \times 10^{-5} 
~\mathrm{Hz}\left(\frac{T_{\rm RH}}{100~\mathrm{GeV}}\right)\left(\frac{g_\star}{100}\right)^{\frac{1}{6}}\,\,,
\end{equation}
where $ \kappa_{\mathrm{turb} }$ represents how effectively vacuum energy is converted into turbulent flow:
	\begin{equation}
		\kappa_{\text {turb }}=\tilde{\epsilon} \kappa_v\,\,,
	\end{equation}
	We set $  \tilde{\epsilon} =0.1 $.
	The peak frequency of turbulence processes is :
	\begin{equation}
		f_{\text {turb }} \simeq 2.7 \times 10^{-5} \mathrm{~Hz}\frac{1}{v_w} \left(\frac{\beta}{H_p}\right)\left(\frac{T_{\rm RH}}{100~ \mathrm{GeV}}\right)\left(\frac{g_\star}{100}\right)^{1 / 6}\,\,.
\end{equation}
\end{itemize}

The total contribution to the SGWB can be calculated by integrating these individual contributions:
\begin{equation}
    h^2 \Omega_{\mathrm{total}}(f)=h^2 \Omega_{\mathrm{co}}(f)+h^2 \Omega_{\mathrm{sw}}(f)+h^2 \Omega_{\mathrm{turb}}(f)\,\,.
\end{equation}


\end{document}